\documentclass{aa}
\usepackage{graphicx}
\usepackage{natbib}
\usepackage{amsmath}
\begin{document}
   \title{Long-term photometric behaviour of XZ~Dra}

   \subtitle{Binarity or magnetic cycle of a Blazhko type RRab star}

   \author{J. Jurcsik, J. M. Benk\H o         
          \and          B. Szeidl
        }


   \institute{Konkoly Observatory of the Hungarian Academy of Sciences. P.O.
Box 67, H-1525 Budapest, Hungary\\
              \email{jurcsik,benko,szeidl@konkoly.hu}
             }

   \date{Received ; accepted }

   \abstract{
The extended photometry available for XZ Dra, a Blazhko type
RR Lyrae star, makes it possible to study its long-term behavior. 
It is shown that its pulsation period  exhibit cyclic, but not
strictly regular variations with $\approx7200$ d period.
The Blazhko period ($\approx76$ d) seems to follow the observed period changes of 
the fundamental mode pulsation with 
${\rm d}P_{\rm B}/{\rm d}P_{0}=7.7\times 10^{4}$ gradient.
Binary model cannot explain this order of period change of the 
Blazhko modulation, nevertheless it can be brought into agreement 
with the $O-C$ data of the pulsation. 
The possibility of occurrence of magnetic cycle is raised.        
 }  

   \maketitle


\section{Introduction}

One of the unsolved problems, which is perhaps the most intriguing 
one in RR Lyrae star research, is the Blazhko effect, the amplitude 
and/or phase modulation of light curves with periods of 
$10 - 500$ days. Although theoretical interpretations of the phenomenon 
have been suggested \citep{shiba,vanh}, a clear explanation of it is 
still lacking.

\citet{blazhko} was the very first who recognized that the light 
maximum of an RR Lyrae star (namely RW Dra) showed phase modulation 
on a long time-scale (around 40 days). Subsequent investigation has 
revealed that the phase modulation is accompanied by modulation of 
the light curve and the amplitude of light variation. About a third 
of the fundamental mode RR Lyrae stars show the effect \citep{szeidl}, 
however, only few have been investigated in detail yet. There are 
less than 10 galactic field stars which have sufficient observations
to permit deeper insight into their Blazhko properties, but these
studies have not been enough to expose the physical background of 
the observed modulation. The number of Blazhko stars which have been 
observed for a long enough time to detect any changes in their 
modulation properties is even fewer. Most of the stars listed in 
the summary review paper of Blazhko variables \citep[completed in
\citealt{smith}]{szeidl} still have not been studied in detail.

XZ Dra 
(BD$+64\degr 1332$, HIP\,94134, $\alpha_{2000}=19^{\rm h} 09^{\rm m}
42\fs6$,
$\delta_{2000}=+64\degr 51\arcmin 32\arcsec$) 
is one of the best observed RRab stars. \citet{schn} discovered the 
star's variability on Babelsberg plates. Soon after the announcement of 
the discovery, \citet{beyer} observed the star visually and determined 
the correct value of the fundamental period. He also commented on the 
strong oscillation in brightness of the individual light maxima.
\citet{bd}, based on the rough estimates of their photographic 
observations showed that these oscillations had a period of 76 days and 
the star's behaviour resembled that of AR Her.

During the last century, continuous effort has been made at the Konkoly 
Observatory to regularly observe RR~Lyrae stars with Blazhko effect. 
Collection of photometric and some radial velocity observations of 
XZ Dra has been recently published in \citet{mitteil}. These data, 
together with all the published measurements of XZ~Dra, made it 
possible to follow its photometric behaviour during a remarkably
long (70-year) period. Due to the extended data now available a detailed 
analysis of the properties of its pulsation and Blazhko behaviour has 
become feasible.

\section{The data}

\begin{figure*}[t]
   \centering
   \includegraphics[width=18.3cm]{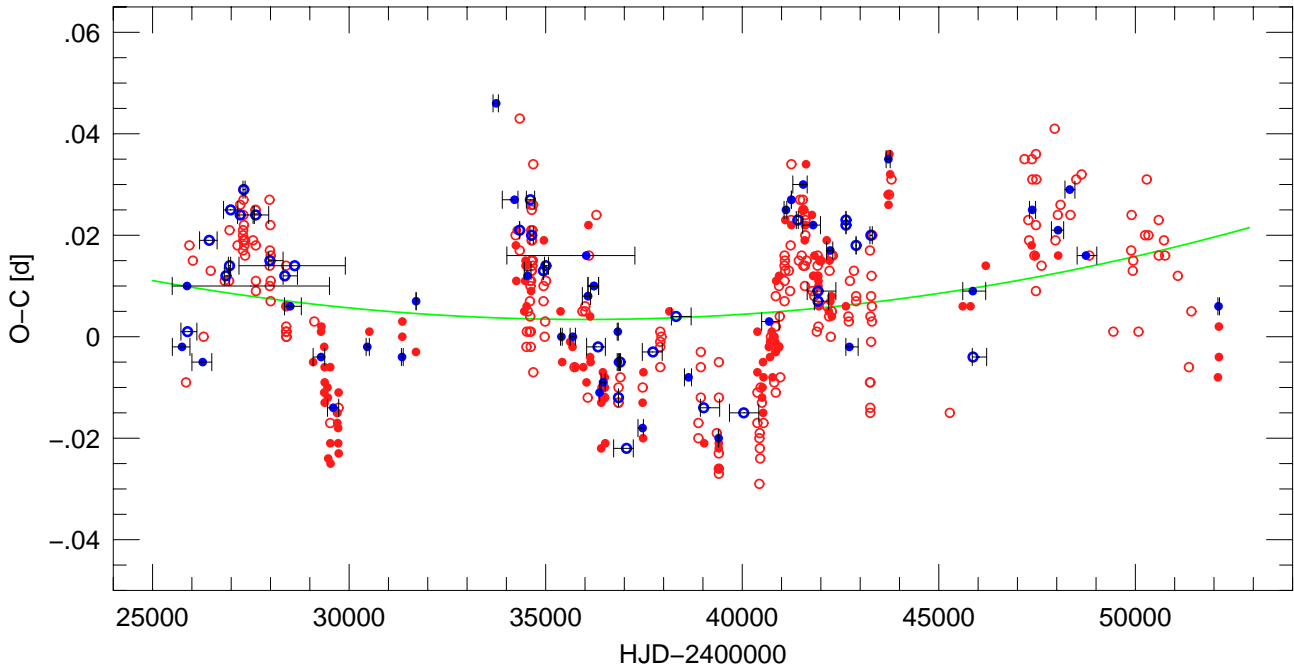}
   \caption{
$O-C$ diagram of XZ~Dra constructed from individual (red) and normal 
(blue) maxima compiled in \cite{mitteil}. Filled symbols are for 
professional photographic, photoelectric and CCD observations. 
Open symbols show visual and amateur data, and also all the uncertain
observations noted by colon in \cite{mitteil}. Horizontal bars 
indicate the time span of the observations the normal maxima were 
derived from. The green line is the least square parabolic fit to 
all the data.}
\vskip 2mm
              \label{oc}
\end{figure*}

\begin{figure}[bhh!!!!!!!!!!!!!!!!]
   \centering
   \includegraphics[width=8.7cm]{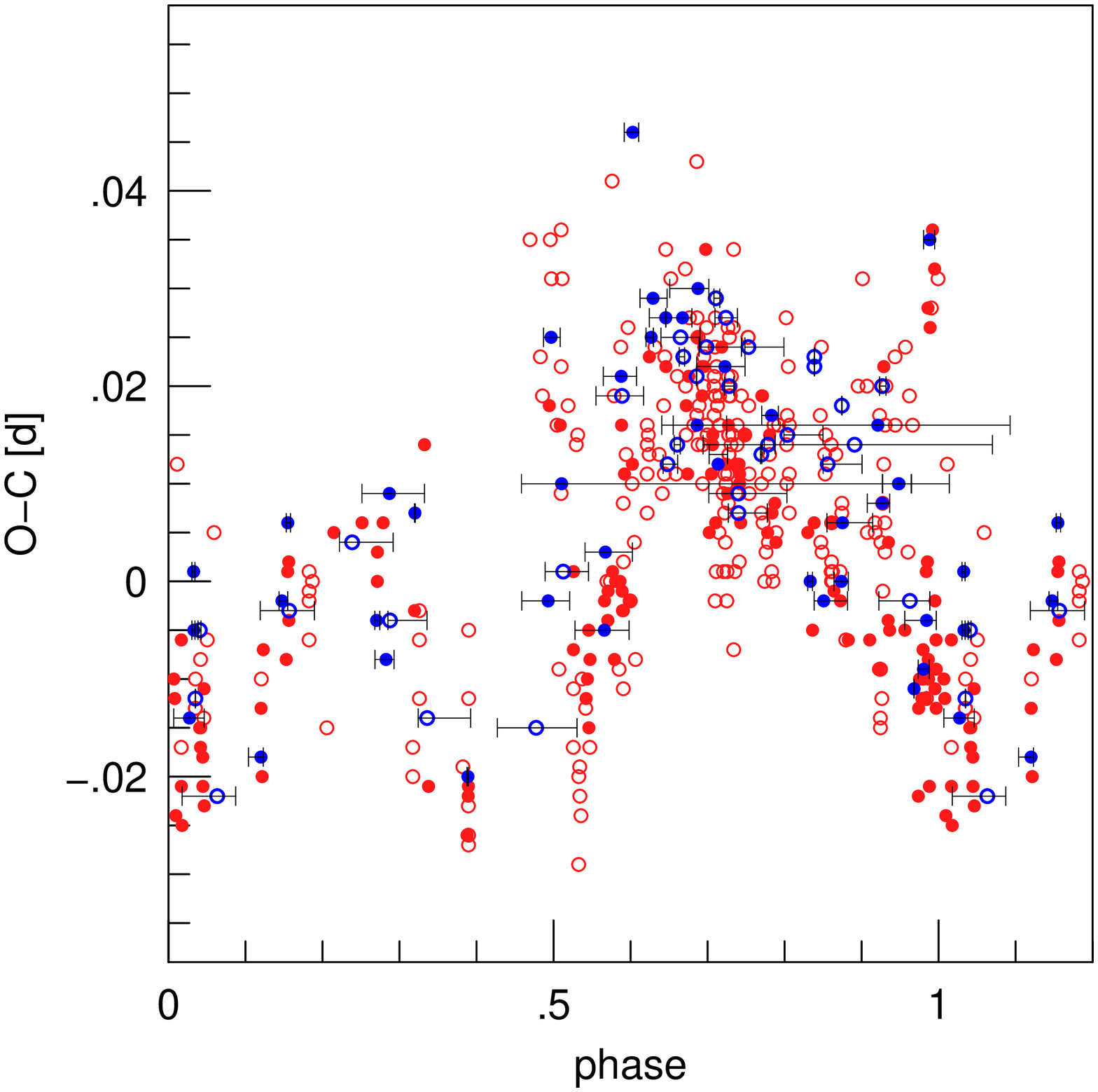}
   \caption{
$O-C$ data folded with $7200$~d. Symbols are the same as in Fig.~\ref{oc},
the parabolic fit to the data has been extracted. The maxima and the 
decline phases of the $O-C$ seem to be closely repetitive, but at around 
minimal values, the long-term changes have a much more irregular character.}
              \label{ocfold}   
\end{figure}

Both the extended photometric and the scarce radial velocity observations 
available for XZ Dra have been utilized in the present study. The photometric 
data used are the photoelectric, CCD, photographic and visual observations 
collected in \cite{mitteil} and also all the other published photometries.
A complete reference list of all the available photometric data of XZ~Dra 
were also given in \cite{mitteil}.

In order to obtain the best possible time coverage of the observations 
the different types of measurements were combined if they could be reliably
transformed to the same magnitude scale and if they belonged to a time 
interval for which data were analyzed together. Thus $B_{\rm pg}$ and 
$V_{\rm pg}$ (photographic) data were transformed to $B_{\rm pe}$ and
$V_{\rm pe}$ (photoelectric) 
magnitudes, respectively, by correcting for zero point offsets. We note, 
however, that only very few and scarce photographic measurements
\citep{zaleski,hard} were transformed in this way, as for most part of the 
photographic data there were no contemporary photoelectric observations.
The visual observations were treated separately, simultaneous visual 
observations were linearly transformed to gain a common magnitude scale. 

The most deviant points were omitted from each of the data sets.

If there were no observations during the descending part of the light 
curve (typically that was the case in the photoelectric data and also 
in some parts of the photographic observations)  $5-15$ artificial 
data were added in order to ensure the stability of the light curve 
solutions. The small number of artificial data compared to the 
total number of the real observations provided that the results were 
not biased by the actual choice of the artificial data.

When analyzing the magnitudes of maxima, all the observations of the 
same wavelength were combined in order to obtain the best possible time 
coverage. For this purpose the $V$ and visual magnitudes of maxima were 
also combined by simply correcting for zero point shifts between the 
$V_{\rm pe}$, $V_{\rm pg}$ and the visual data. 

A complete list of individual and normal maximum timings
(an average maximum timing determined from observations of more than 
one pulsation cycle taking into account the whole light curve as well 
not only times of maxima) and $O-C$ values using the ephemeris
\begin{equation}
{\rm Max(HJD)}=2\,431\,244.383+ 0.4764955\, E
\end{equation}
were also given in Table 10.a and 10.b of \citet{mitteil}. 
In the present paper these $O-C$ data are also elaborated. 

Although only very limited radial velocity observations of XZ Dra are 
available, these observations are also reviewed (Sect. 4.3) in comparison
with the possible explanations of the photometric data.

\section{Period changes}

The collection of all the individual and normal maximum timings of 
XZ~Dra allows us to follow its period changes using a nearly continuous 
data base covering 70 years. The $O-C$ diagram constructed from 
this compilation is shown in Fig.~\ref{oc}. According to Fig.~\ref{oc} 
the mean pulsation period of XZ~Dra has not changed significantly 
during the time covered by observations, but $O-C$ variations ranging 
$0.02-0.04$~d in $2000-5000$~d intervals indicate that opposite sign 
period changes of the order of $10^{-5}$~d/d have been occurring.

A detailed inspection of the $O-C$ data exposed that besides a small 
continuous period increase, a $\approx7200$~d cyclic variation can be 
also detected. In Section 3.1 these long-term changes are documented. 
Thorough analysis of all the available photometric observations have been 
performed in Section 3.2, with a special focus on  the possible 
changes in the Blazhko properties consequent to the detected variations 
of the pulsation period.

\subsection{Long-term and sudden period changes}

Fourier analysis of the $O-C$ data shown in Fig.~\ref{oc} revealed 
that the period change behaviour of XZ~Dra can be described with a 
$\approx7200$~d asymmetric shape periodic modulation. Filtering out this 
cycle, it becomes evident that a slight, continuous period increase has 
also occurred, as the residual $O-C$ clearly has a parabolic shape. 
In Fig.~\ref{oc} the parabolic fit to the data is also shown. 
According to the parameters of the quadratic fit the ephemeris 
of XZ~Dra has been modified to:
\begin{equation}
\begin{split}
{\rm Max(HJD)}=2\,435\,819.2190(9)+ 0.47649550(6)\, E+\\
1.9(3)\times10^{-11}\, E^{2}
\end{split}
\end{equation}

The continuous period change corresponds to $2\times10^{-6}$~d period 
increase during the 25\,000 days covered by the observations.
As this period change is negligible if compared to the much more 
effective, shorter time-scale period changes, in the course 
of the analysis performed in Section 3.2 this continuous period 
increase has been ignored.

In order to document the cyclic nature of the long-term variability 
seen in Fig.~\ref{oc}, the $O-C$ data, after removing the parabolic 
fit, is folded with 7200~d in Fig.~\ref{ocfold}. The best period found 
by fitting a sinusoid with 2 harmonics to the data is $7198\pm100$~d. 
The maxima and the descending
parts of the $O-C$ data follow this periodicity well, but the larger 
scatter of the folded curve around minimum and ascending branch 
indicates that there are differences in the shape of the $O-C$ changes
from one cycle to the other.

It is also important to note that besides the parabolic and the 7200~d 
cyclic variations of the  $O-C$ there are also sudden changes observed, 
see e.g. the $0.02$~d increase around HJD~2\,443\,800.

\subsection{Pulsation and Blazhko periods}

The observed changes in the pulsation period raise the issue of the 
concurrent behaviour of the Blazhko modulation. An obvious way to 
follow the possible variations of the Blazhko period is to construct 
the $O-C$ diagram corresponding to the Blazhko periodicity. Such an 
attempt has, however, failed. The time coverage of the observations, and 
the not strictly regular nature of the Blazhko modulation, in most of
the cases permit only an estimation of the observed times of maxima of the 
Blazhko cycles with $5-15$ days accuracy. To draw firm conclusions 
about smaller period changes this accuracy is not enough. We thus 
mention only that all the $O-C$ values of the maximal phases of the 
Blazhko period spread within a 15 days range if calculated using a 
75.7~d mean Blazhko period value. 

We studied the simultaneous properties of the two periodicities 
using all the available photometric observations. The data were not 
corrected for the continuous period change shown in Sec. 3.1, 
because of its negligible effect along the time intervals of the 
analyzed data sets.

First, the data were divided into different segments according to the 
structure of the $O-C$ diagram. Linear parts of the $O-C$ were selected, 
assuming that along these segments no significant period change has
been occurring. From different trial divisions of the data it turned out, 
that the folded light curves of those data sets which could definitely 
be described with a constant pulsation period value, always had a 
fix point in their rising branch. This behaviour of the modulation was 
also a guide in selecting data groups, considered not being biased 
by period change that mimics phase modulation. 

Both the full light curves (Section 3.2.1) and the data of maximum
timings and magnitudes (Section 3.2.3) were analyzed.

\begin{figure}[tttttt!!!]
   \centering
   \includegraphics[width=9cm]{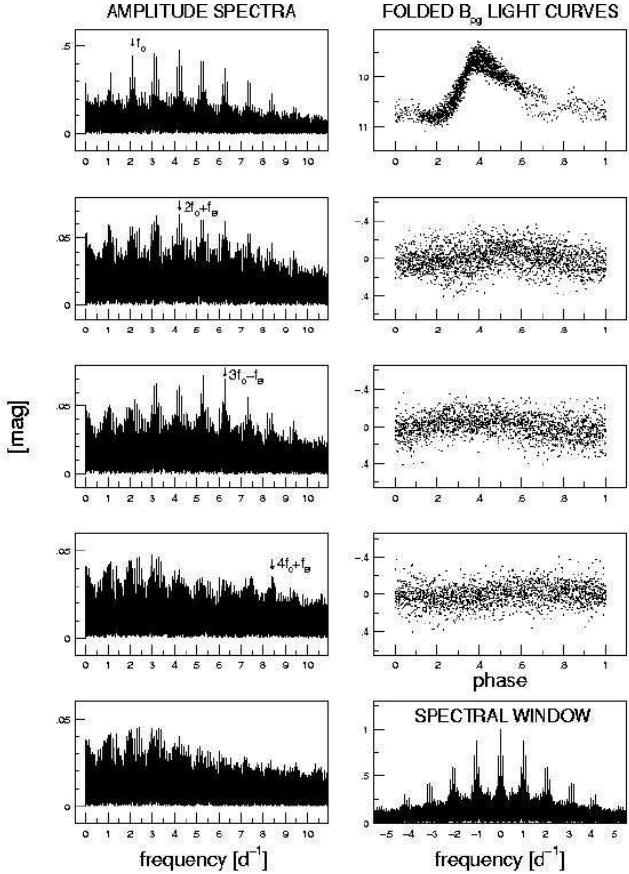}
   \caption{
Fourier spectra and folded light curves of the photographic observations 
between JD~2\,429\,084 and JD~2\,429\,734 in the course of prewhitening 
first with the fundamental mode period and its harmonics then with 
frequencies corresponding to one of the possible modulation frequencies
appearing in the consequent prewhitened spectra ($2f_0+f_{\rm B}$, 
$3f_0-f_{\rm B}$, $4f_0+f_{\rm B}$). These frequencies are indicated by 
arrows in the spectra and the residual light curves are folded according 
to these periods.}
              \label{sp}
\end{figure}

\begin{table}[bbb!!]
\caption{Trial modulation frequency combinations used to fit the 
Hipparcos \citep{hip} (H1, H2 fits) and \citet{dyach} (D1, D2 fits) 
data. Besides the modulation frequencies listed, 5 and 4 harmonics
of the pulsation frequency were also taken into account in the 
H1, H2 and D1, D2 fits, respectively. The $P_0$, $P_{\rm B}$ periods 
correspond to the minimum of the two dimensional {\it rms} distributions 
(see also Figs.~\ref{hip} and Fig.~\ref{dats}).
\vskip -7mm
}
         \label{amp}
      $$
         \begin{tabular}{lcccc}
            \hline
            \noalign{\smallskip}
                  & H1 fit & H2 fit & D1 fit & D2 fit \\
           \noalign{\smallskip}
            \hline
           \noalign{\smallskip}
$P_0$ [d]            & 0.4764950& 0.4764950& 0.4765035& 0.4765030\\
$P_{\rm B}$ [d]          & 75.7     & 76.1     & 77.9     & 78.0\\
           \noalign{\smallskip}
            \hline
           \noalign{\smallskip}
\multicolumn{1}{c}{mod. fr.}  & \multicolumn{4}{c}{Amplitudes [mag]}\\ 
           \noalign{\smallskip}
            \hline
           \noalign{\smallskip}
$f_0-f_{\rm B}$ &0.0239&0.0255&0.0750&0.0676\\
$f_0+f_{\rm B}$ &0.0338&0.0311&0.0036&   $-$\\
$2f_0-f_{\rm B}$&0.0243& $-$  &0.0622&0.0364\\
$2f_0+f_{\rm B}$&0.0200&0.0226&0.0298&  $-$\\
$3f_0-f_{\rm B}$&0.0088&  $-$ &   $-$&0.0551\\
$3f_0+f_{\rm B}$&0.0132&0.0240&  $-$ &$-$ \\
$4f_0-f_{\rm B}$& $-$  & $-$  &  $-$ &$-$ \\
$4f_0+f_{\rm B}$& $-$  &0.0317&   $-$&$-$ \\
$f_{\rm B}$     & $-$  &0.0268&  $-$ &$-$ \\
      \noalign{\smallskip}
            \hline
         \end{tabular}
$$
\end{table}

\begin{figure}[tttt]
   \centering
   \includegraphics[width=9cm]{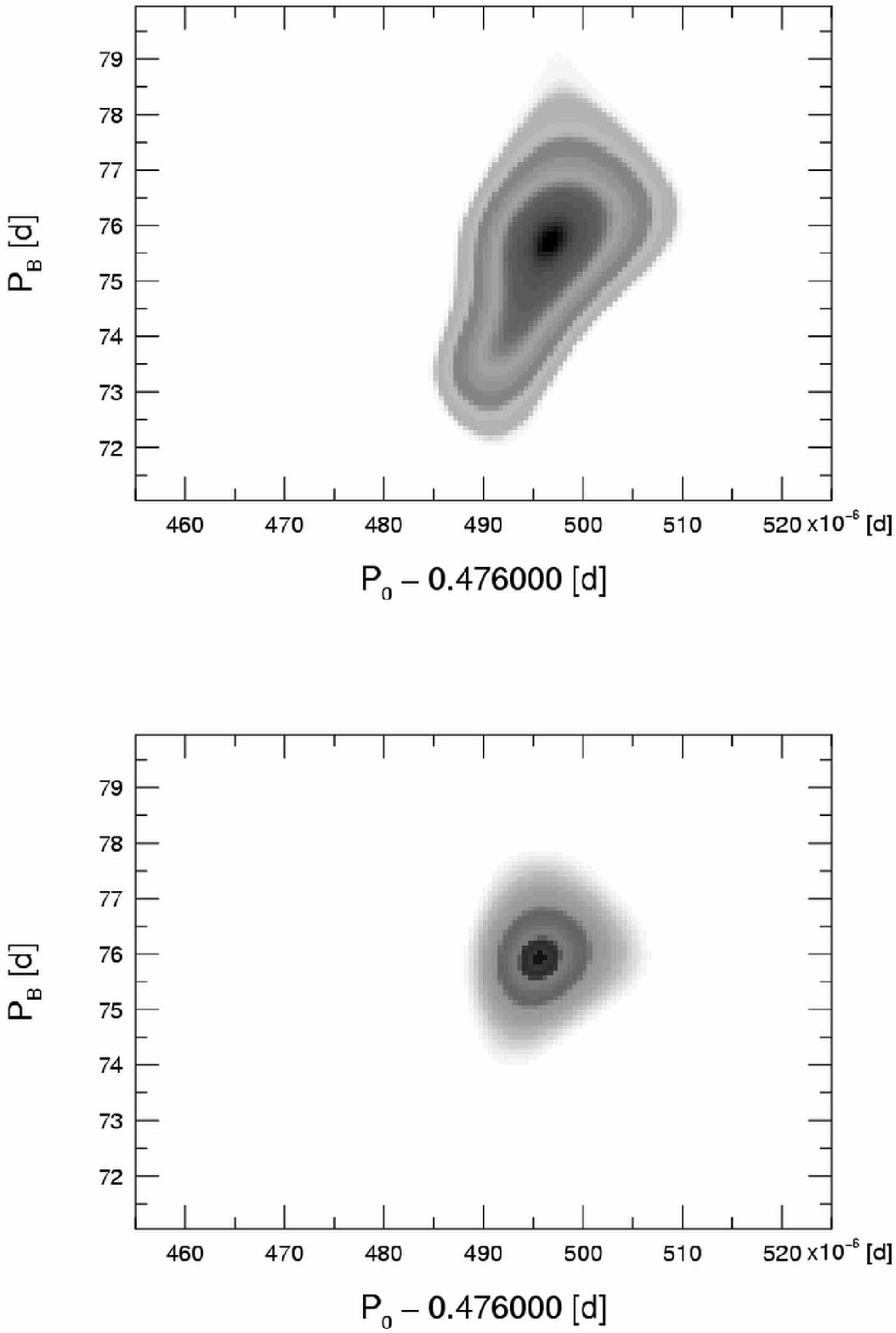}
   \caption{
The {\it rms} scatter of the Hipparcos data on an arbitrary gray
scale. Fits were calculated using pulsation and Blazhko periods
indicated in the axes. The unit of the horizontal scale is $10^{-6}$ d.  
The pulsation frequency with 5 harmonics and modulation frequencies
given in Table~\ref{amp} were considered when determining the {\it rms} 
scatter of the fits. Solutions using modulation frequencies listed in 
Table~\ref{amp} for H1~fit are shown in the upper, for H2~fit in the 
lower panels, respectively. The {\it rms} range shown is the increment 
of the minimal {\it rms} value by $\approx10$\%.}
\label{hip}
\end{figure}

\begin{figure}[ttttt]
   \centering
   \includegraphics[width=9cm]{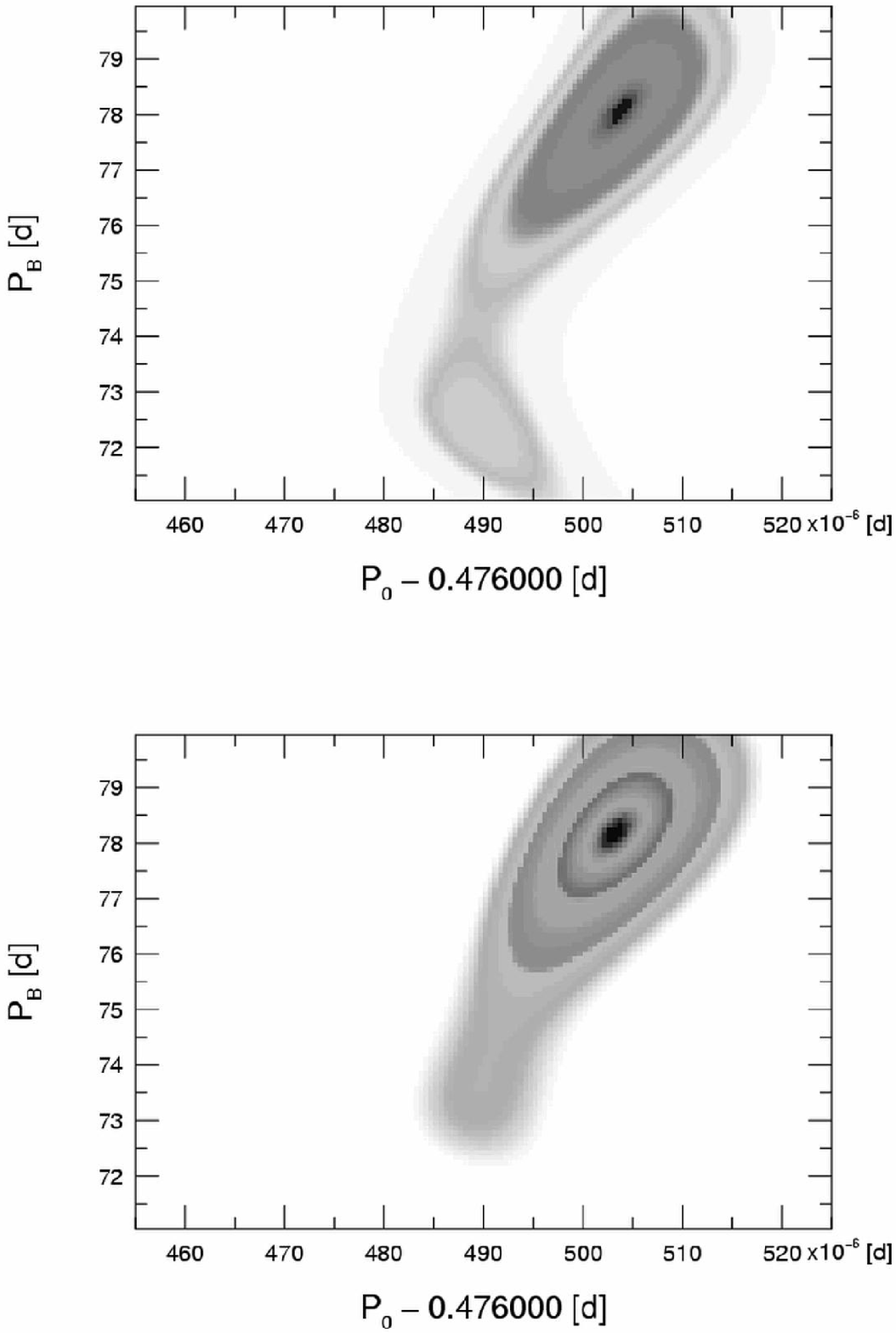}
   \caption{
The same as in Fig.~\ref{hip} but for \citet{dyach} visual observations. 
Solutions for D1 and D2 fits (Table~\ref{amp}) are shown in the upper and 
lower panels, respectively. $0.086-0.090$~mag {\it rms} range is shown in 
both panels.}
   \label{dats}
\end{figure}

\subsubsection{Analyzing light curves}

Different methods were used to determine the pulsation and Blazhko 
periods valid for the individual data sets. The governing principle 
was to find $f_0$ (fundamental mode frequency) and $f_{\rm B}$ 
(Blazhko frequency) that gave the best fit using $f_0$ and $4-8$ of 
its harmonics and $3-9$ additional frequencies among the modulation 
frequency pattern commonly found in the spectra of Blazhko variables: 
$k\times f_0 \pm f_{\rm B}$ (k=1,2,3,4) and $f_{\rm B}$.
Fourier decomposition using the MUFRAN program package \citep{mufran},
least square minimization technique and nonlinear regression facilities 
of {\sc Mathematica}\footnotemark[1] \citep{Mat}\footnotetext[1]{{\sc 
Mathematica} is a registered trademark of Wolfram Research Inc.} 
have been used in deciding which modulation frequencies to take into 
account for the individual data sets, and to determine the actually 
valid $f_0$ and $f_{\rm B}$ frequencies.

In many cases it was not possible, however, to decide which modulation 
frequencies to take into account, as the shortness of the data set 
and/or its high noise, limited the parameter number of the solutions. 
In these cases many possibilities were tested and $f_0$ and $f_{\rm B}$ 
were only accepted if the same periods consistently appeared in the 
different solutions. Results with unreliable large amplitudes of the 
modulation frequencies were rejected.

\begin{figure}[tttthhh]
   \centering
   \includegraphics[width=9cm]{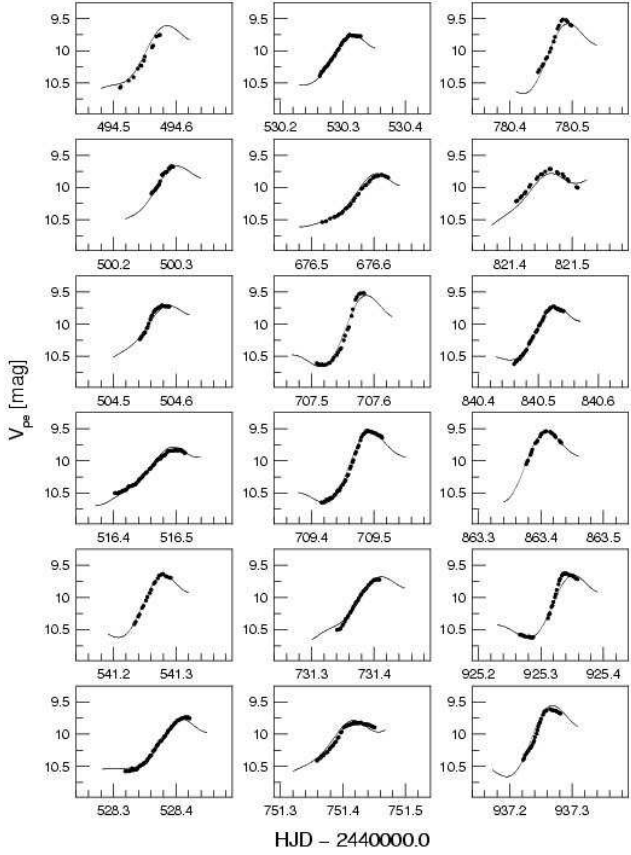}
   \caption{
Photoelectric $V_{\rm pe}$ light curves fitted with $f_0$ (and 6 of its harmonics), 
$2f_0+f_{\rm B}$, $3f_0+f_{\rm B}$, $4f_0-f_{\rm B}$ frequencies using 
$P_0=0.476516$~d and $P_{\rm B}=76.69$~d determined from nonlinear
regression to the data.}
              \label{fit}
\end{figure}

Some examples of the applied procedure are documented in
Figs.~\ref{sp},~\ref{hip},~\ref{dats},~\ref{fit},  showing results for 
different data sets. 

Due to the shortness and inconvenience of the data sampling of most of the data
sets, their Fourier spectra are usually crowded by alias peaks which make
it difficult to identify the real frequencies. At least one of the possible
modulation frequencies, however, can be always detected in the spectra.
Interestingly, this dominant modulation frequency is usually found near
the higher harmonics ($2-4$) of the fundamental mode frequency. We do not
know at present whether this is a real feature of the modulation, or this
arises from the bias of the unfavourable data distribution.

The Fourier analysis of some data sets with good phase coverage can be
also misleading in some cases. For example, the Fourier spectrum of the
photographic data between JD~2\,430\,433 and JD~2\,431\,708 shows modulation
frequencies at 0.0187 and 0.0243 c/d (53.5 and 41.2 d) distant from the
harmonics of the pulsation frequency, which might be aliases of the 
true modulation peaks. Notwithstanding this, least square fit
using 8 symmetrically placed modulation frequencies indicates better,
more reliable solutions at 0.01302 and 0.01328 c/d (76.8~and~75.3~d).

A sample Fourier spectra of the photographic observations between
JD~2\,429\,084 and JD~2\,429\,734 is shown in Fig.~\ref{sp}. The procedure
of first prewhitening with  the fundamental mode period (with 6 harmonics),
then with the modulation frequencies appearing in the subsequent whitened
spectra is documented in the left panels. The right panels show the folded
light curves using the periods indicated by arrows to the corresponding
spectra.

Fig.~\ref{hip} and Fig.~\ref{dats} show the two 
dimensional residual scatter of the Hipparcos \citep{hip} and \citet{dyach} 
light curves fitted by two different `Blazhko type' frequency patterns 
listed in Table~\ref{amp}. For each possible $P_{0}$ and $P_{\rm B}$ 
pairs the {\it rms} residuals are shown in an arbitrary gray scale 
to set off the {\it rms} distribution in the vicinity of the best solution. 
It can be seen in Fig.~\ref{hip} and Fig.~\ref{dats} that the structure of 
the minimum places changes significantly for the different frequency 
solutions of a given data set but the location of the absolute minimum 
place remains very closely at the same $P_{0}$ and $P_{\rm B}$ period pair. 
Table~\ref{amp} lists the modulation frequencies that were taken into 
account when calculating the {\it rms} of the fits shown in Figs.~\ref{hip},
~\ref{dats}, and their amplitudes at $P_{0}$ and $P_{\rm B}$ periods 
of the absolute minimum places. It is worth noting that the amplitudes of 
the same modulation frequency differ by as much as $30-50\%$ if different 
frequency patterns are used. This result indicates, that {\it any speculation 
concerning possible amplitude changes of a given modulation frequency 
has to be taken with caution.}

\begin{table}[bbb!!!!]
\caption{Pulsation and Blazhko periods derived from the light curves}
         \label{per0}
      $$
         \begin{tabular}{r@{\hspace{9pt}}l@{\hspace{4pt}}l@{\hspace{6pt}}l@{\hspace{3pt}}l}
            \hline
            \noalign{\smallskip}
\multicolumn{1}{c}{JD}      &  $P_0$\hskip13pt error&   $P_{\rm
B}$\,\, error &
source&refs.\\
            \noalign{\smallskip}
\multicolumn{1}{c}{$-2\,400\,000$}  &  [d]\,\,\,\, $10^{-6}$[d]& \multicolumn{1}{c}{
[d] }
&& \\
           \noalign{\smallskip}
            \hline
           \noalign{\smallskip}
$25721 - 27556$&	0.476502 2	&76.77	0.43& vis.       &1,2,3\\
$28356 - 29141$&	0.476491 1	&76.19	0.23& $B_{\rm pg}$ &1\\
$29084 - 29734$&	0.476485 1	&74.74	0.35& $B_{\rm pg}$     &1\\
$^{\mathrm{*}}30433 - 31708$&	0.476495 2	&75.27	0.30& $B_{\rm pg}$ &1 \\
               &	0.476495 2	&76.80	0.25&              &  \\
$33661 - 34627$&	0.476462 4	&73.73	0.55& $B_{\rm pg}$ &1,4\\
$34335 - 35072$&	0.476483 3	&73.00	1.40& vis.       &5,6\\
$35377 - 36347$&	0.476499 2	&76.64	0.37& $B_{\rm pg}$ &1,4\\	
$36041 - 37583$&	0.476491 3	&76.13	0.59& vis.       &7\\	
$36410 - 39403$&	0.476493 1	&75.79	0.05& $V_{\rm pe}$,
$V_{\rm pg}$&1,4,8 \\	
$36410 - 39403$&	0.476493 0.5	&75.80	0.05& $B_{\rm pe}$,
$B_{\rm pg}$&1,4,8,9 \\	
$37844 - 40121$&	0.476487 3	&75.59	0.65& vis.       &7\\	
$40494 - 40937$&	0.476515 1	&76.75	0.42& $V_{\rm pe}$ &1\\	
$40494 - 40937$&	0.476514 1	&76.40	0.45& $B_{\rm pe}$ &1\\	
$41249 - 42636$&	0.476489 1	&75.60	0.14& $V_{\rm pe}$ &1\\	
$41249 - 42636$&	0.476489 1	&75.61	0.15& $B_{\rm pe}$ &1\\
$42894 - 43308$&	0.476503 5	&78.04	1.13& vis.       &10\\	
$47865 - 49019$&	0.476496 3	&75.90	1.25& $V_{\rm Hip.}$    &11\\
            \noalign{\smallskip}
            \hline
         \end{tabular}
     $$ 
{\footnotesize
\underline {References:}
1 \citet{mitteil},
2 \citet{beyer},
3 \citet{lange},
4 \citet{zaleski},
5 \citet{batyr},
6 \citet{klep},
7 \citet{lebed},
8 \citet{sturch},
9 \citet{hard},
10 \citet{dyach},
11 \citet{hip}}
\begin{list}{}{}
\item[$^{\mathrm{*}}$] Two Blazhko periods are equally possible
\end{list}
\end{table}

The photoelectric observations seriously lack measurements taken during the 
descending branch of the light curves. It makes direct Fourier analysis 
almost impossible, but nonlinear regression assuming appropriate frequencies 
can still yield reliable solution for $P_0$ and $P_{\rm B}$. For this 
purpose the nonlinear regression facilities of {\sc Mathematica} assuming 
trial modulation frequency combinations were used. Fig~\ref{fit} demonstrates
such a result, photoelectric $V$ observations of the JD~$2\,440\,494-2\,440\,937$
period are fitted with $f_0$, $2f_0+f_{\rm B}$, $3f_0+f_{\rm B}$, and 
$4f_0-f_{\rm B}$, that was found, after many trials using different possible
modulation frequencies, as a reliable solution.

Periods obtained during the course of the above procedure for the different 
time intervals (supposed not to be affected by period changes), are summarized 
in Table~\ref{per0}. The periods determined for the time intervals given in 
Column~1 are the average values of different solutions, while estimating 
their errors both the range of the period values obtained and
their individual errors are taken into account.

\subsubsection{Periods determined from maximum timings and magnitudes}

In Table~\ref{per1} periods determined from maximum timings and magnitudes 
are listed. $V$ band ($V_{\rm pe}$, $V_{\rm pg}$ and visual after 
transformed to approximately match the $V_{\rm pe}$ magnitudes), 
$B$ band ($B_{\rm pe}$, $B_{\rm pg}$) maximum magnitudes and $B-V$ 
photoelectric observations of maxima were used independently. The $B-V$ maxima 
were determined from the Konkoly photoelectric $B$ and $V$ observations. 
$B-V$ curves were constructed by extrapolating the observations to the same 
instants, and the maxima of the colour curves were determined instead of 
taking $B_{\rm max}-V_{\rm max}$ values. 

These data were also divided into segments corresponding to linear parts 
of the $O-C$ curve, however, the time intervals selected were not exactly
identical to those chosen in Sect. 3.2.1.

The Blazhko periods were determined from single sinusoidal fits to the 
maximum magnitude values while the simultaneous pulsation periods were 
obtained from linear fits to the maximum timings, {\it i.e.} the $O-C$ 
values \citep[][Table 10.a]{mitteil} of the time interval considered. 
From the slope of the $O-C$ curve it was determined what the instantaneous 
pulsation period actually was.

Table~\ref{per1} lists $P_0$ and $P_{\rm B}$ for the different time intervals
determined this way. Cols.~$1-3$ give the starting and ending dates of the 
$O-C$ data the pulsation period was calculated from, the $P_0$ period obtained, 
and its error. Cols.~$4-7$, Cols.~$8-11$, and Cols.~$12-15$ list similar data 
for the Blazhko period, determined from $B$, $V$, and $B-V$ maximum magnitudes, 
and also references of the photometries used.
\begin{table*}[ttt!!!]
\caption{Pulsation and Blazhko periods derived from times and magnitudes of maxima}
         \label{per1}
$$
         \begin{tabular}{
l@{\hspace{6pt}}l@{\hspace{6pt}}l@{\hspace{5pt}}l@{\hspace{5pt}}l@{\hspace{5pt}}l@{\hspace{5pt}}l@{\hspace{5pt}}l@{\hspace{5pt}}l@{\hspace{5pt}}l@{\hspace{5pt}}l}
            \hline
            \noalign{\smallskip}
\multicolumn{1}{c}{JD}&  $P_0$ \hskip 12pt error& \multicolumn{1}{c}{JD}
& $P_{\rm B}^{\mathrm{a}}$ \hskip 8pt error & refs.&\multicolumn{1}{c}{JD}& $P_{\rm
B}^{\mathrm{b}}$\hskip 8pt error &
refs.&\multicolumn{1}{c}{JD}& $P_{\rm B}^{\mathrm{c}}$ \hskip 8pt error & refs.\\
            \noalign{\smallskip}
\multicolumn{1}{c}{$-2\,400\,000$}& [d]\,\,\,\,$10^{-6}$[d]&
\multicolumn{1}{c}{$-2\,400\,000$}& \multicolumn{1}{c}{
[d] } &&\multicolumn{1}{c}{$-2\,400\,000$} 
                  &\multicolumn{1}{c}{  [d] } &&\multicolumn{1}{c}{$-2\,400\,000$}
                  &\multicolumn{1}{c}{  [d] } &\\
           \noalign{\smallskip}
\hline
           \noalign{\smallskip}
$25850-27556$& 0.476501 1&&&& $25850-27556$&77.46 1.03 &1,2,3 & &&\\	
$27556-28409$& 0.476486 2&&&& $27618-28409$&75.03 1.49 &4 & &&\\		
$29084-29734$& 0.476482 3& $29084-29734$&73.52 0.83 &1 &&\\	
$33740-34627$& 0.476473 4& $34234-34627$&73.03 2.10 &1 & &&&&&\\
$35377-36289$& 0.476502 4& $35377-36289$&77.18 2.31 &1,5 & &&&&&\\
$36410-39406$& 0.476495 1& $36410-39403$&75.82 0.20 &1,5,6 & $36410-39403$ &75.95 0.20 &1,5,6& & &\\	
$40434-40966$& 0.476516 3& $40500-40937$&77.26 0.90 &1 & $40434-40966$ &76.30 0.98 &1,7  
&$40500-40937$&78.10 1.80 &1 \\	
$41187-42739$& 0.476489	1& $41249-42636$&75.78 0.49 &1 & $41187-42739$ &75.28 0.33 &1,7 	&$41249-42636$&76.43 1.56 &1 \\	
$42840-43795$& 0.476512	2&&&& $42895-43763$&75.88 2.22 &1,8 & &&\\	
            \noalign{\smallskip}
            \hline
         \end{tabular}
     $$ 
{\footnotesize
\underline {References:}
1 \citet{mitteil},
2 \citet{beyer},
3 \citet{lange},
4 \citet{klep},
5 \citet{zaleski},
6 \citet{fitch},
7 \citet{wenske},
8 \citet{dyach}}
\begin{list}{}{}
\item[$^{\mathrm{a}}$] $P_{\rm B}$ determined from combined $B_{\rm pg}$, $B_{\rm pe}$
maximum magnitudes
\item[$^{\mathrm{b}}$] $P_{\rm B}$ determined from combined $V_{\rm pg}$, $V_{\rm pe}$,
visual maximum magnitudes
\item[$^{\mathrm{c}}$] $P_{\rm B}$ determined from  $(B-V)_{\rm pe}$ maximum magnitudes
\end{list}
   \end{table*}
\begin{figure*}[ttttttttt!!!!!!!!!!!!] 
\vskip -1cm
   \centering
   \includegraphics[width=16cm]{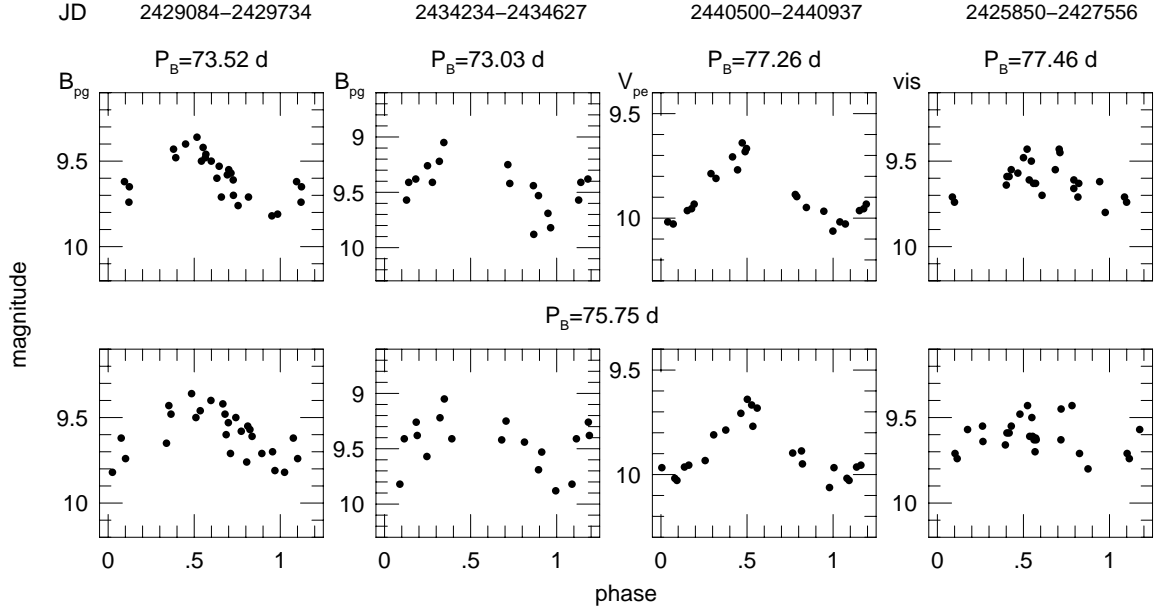}
\vskip -2cm
   \caption{
Maximum magnitudes folded with the best Blazhko periods found
for the different data sets (top panels) and also with the mean 75.75 d Blazhko period
value (bottom panels).
Data sets with the  shortest and longest  Blazhko periods are shown.  
The $rms$ scatter of sine fits of the bottom plots are $20-30$\% larger 
than in the case of the best solutions.
}
              \label{blfold}
\end{figure*}

To demonstrate the reality of changes in the length of the Blazhko cycle
the data sets of light maximum magnitudes corresponding to the shortest and longest
modulation periods as given in Table~\ref{per1} are shown in Fig.~\ref{blfold}
folded according to the mean Blazhko period value (75.75 d) and with the best $P_{\rm B}$
found to be valid for the individual data sets.
In each cases the $rms$ scatter of a sine fit to the data is $20-30$\%
larger if folded with the mean $P_{\rm B}$ value.

\subsubsection{Relation between $P_0$ and $P_{\rm B}$}

\begin{figure}
   \centering
   \includegraphics[width=9cm]{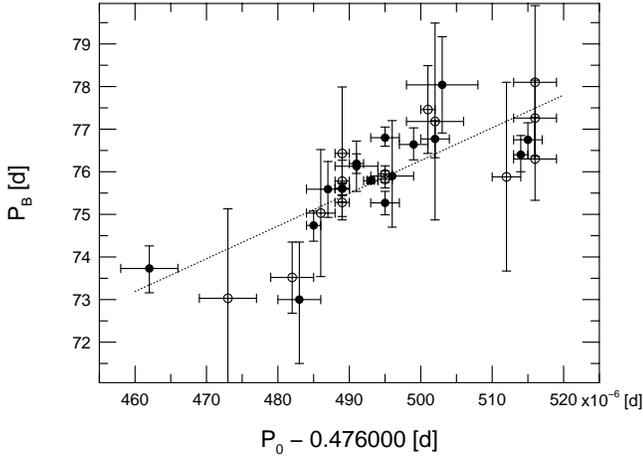}
   \caption{
Pulsation ($P_0$) {\it vs.} Blazhko ($P_{\rm B}$) periods of XZ~Dra. Filled 
and open symbols represent data of Table~\ref{per0} and Table~\ref{per1}, 
respectively. The linear fit to the data has a $7.68\times 10^4$~d/d gradient.}
              \label{per}
\end{figure}

The Blazhko periods as a function of the pulsation periods determined for
the different linear segments of the $O-C$ are shown in Fig.~\ref{per}. 
Filled symbols denote results obtained from analyzing the light curves 
(Table~\ref{per0}) and open circles show the periods determined from maximum 
timings and magnitudes (Table~\ref{per1}). The large uncertainties of the 
periods are mostly due either to the unfavourable data sampling or the 
shortness of the analyzed data sets. The length of the linear parts of 
the $O-C$, however, seriously delimit the length of the data sets used to 
determine the actual period values.

Both sets of data plotted in Fig.~\ref{per} indicate a linear correlation 
between the two periodicities. A least square linear fit to the combined 
data gives a $7.7\pm1.1\times 10^4$~d/d slope of the regression line.

The fact that not only data around the extreme $P_0$ values indicate the 
correlation but the `mid points' also follow a definite trend, strengthens
the validity of the result, namely that in the case of XZ Dra,
{\it $P_0$ and $P_{\rm B}$ periods are not independent quantities but exhibit 
parallel changes}.

Similar connected changes of the pulsation and Blazhko periods have 
been already suggested for XZ Cyg \citep{laclxz}, RW Dra \citep{firm},
and RV UMa \citep{kanyorv}. In these three stars, however, instead of 
parallel, reverse changes of the two periods are suggested.

Possible explanations of the phenomenon are given in Section 5. 

\section{The Blazhko variations}

\subsection{Stability of the Blazhko phase}

\begin{figure}[t]
   \centering
   \includegraphics[width=10cm]{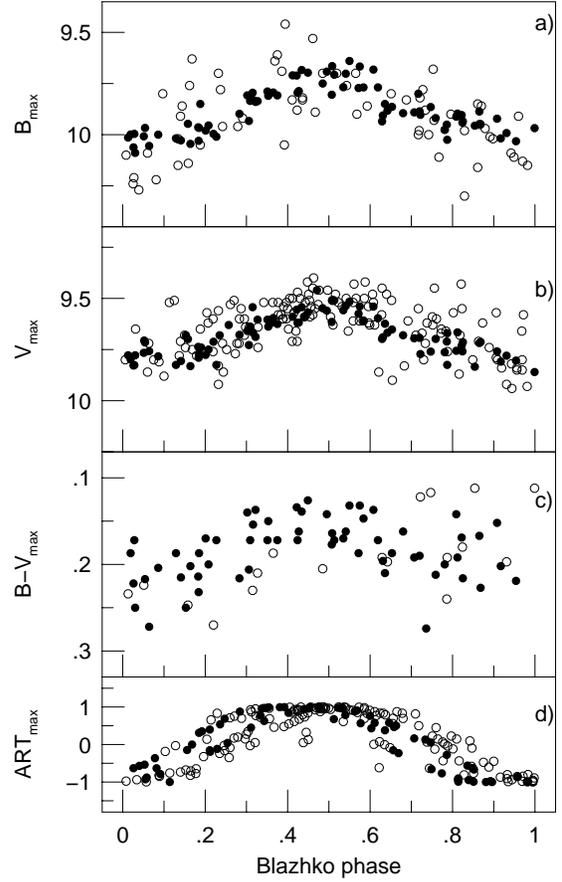}
   \caption{
Folded curves of maxima magnitudes using P=75.75 d mean Blazhko period. 
Panel a) shows $B_{\rm pe}$ (filled circles) and $B_{\rm pg}+0\fm4$ 
(open circles) data of the JD $2\,428\,404 - 2\,447\,462$ time interval. 
Panel b) shows $V_{\rm vis,pg,pe,CCD}$ maxima magnitudes transformed to 
the same magnitude scale (JD $2\,425\,850 - 2\,452\,100$), professional 
photoelectric and CCD observations are denoted by filled circles.
c) In the panel showing $(B-V)_{\rm max}$ magnitudes only photoelectric 
observations are plotted, the less certain values obtained preceding 
JD 2\,438\,000 \citet{mitteil} are denoted by open circles.
Panel d) shows artificial maximum magnitudes which were generated by 
taking into account the observed Blazhko period changes.
Each of the folded curves show relatively small scatter that indicates
phase stability of the Blazhko modulation along the entire time interval
covered by the observations.}
              \label{fold}
\end{figure}

The best studied Blazhko variable, RR Lyrae, shows  pronounced cyclic 
behaviour of its Blazhko modulation. It can be described as a lowering 
of the amplitude of the modulation and a jump in the Blazhko phase in 
about every four years \citep{detre}. On the contrary, XZ Dra shows 
surprising stability of the Blazhko phases during the entire time interval 
covered by observations, notwithstanding the changes of its Blazhko period.

In Fig.~\ref{fold} all the observed maxima ($B_{\rm pg}$ and $B_{\rm pe}$
between JD $2\,428\,404 - 2\,447\,462$, and visual and $V_{\rm pe}$ 
between JD $2\,425\,850 - 2\,452\,100$) are shown. The $B-V$ maximum 
magnitudes are from photoelectric observations. The fact, that all these 
maximum magnitudes can be folded with a common Blazhko period indicates 
strong stability of the Blazhko phases.
 
The relatively small scatter of the folded curves of maximum magnitudes
seems to contradict the detected changes of the Blazhko period (see section 3).
In order to examine the phase incoherence of the maximum magnitudes induced by
the detected Blazhko period changes, artificial data ($\rm{ART_{max}}$) were 
generated and treated similarly to true observations. Using the Blazhko 
periods listed in Table~\ref{per0} and Table~\ref{per1} for the different 
time intervals, artificial maximum magnitudes were generated for the 
JD $2\,425\,850-2\,443\,300$ interval supposing sinusoidal changes with 
varying period. Data sampling follows the distribution of the real observations 
and an arbitrary magnitude scale is used.

Although it is somewhat ambiguous which periods to use to fill the gaps in 
the observations, the simulation shows that the detected period changes do 
not give rise to large divergence in the Blazhko phases. The artificial 
maximum magnitudes, folded with the same Blazhko period as the measured 
instants of maxima, are shown in the bottom panel of Fig.~\ref{fold}.
The scatter of the folded $\rm{ART_{max}}$ curve seems to be similar to 
that of the true observations, thus we can conclude that the phase 
stability does not contradict the period changes of the Blazhko modulation. 
We also recall that the light curve of the pulsation shows similar 
behaviour. The influence of the detected changes of the pulsation period  
on the folded light curve of the entire set of observations is not larger 
than phase shifts of about  $10-15\%$ of the period, which means that 
all the photometric data can be folded with a mean pulsation period 
without larger phase incoherences.

\subsection{Colour changes}

The colour changes of  Blazhko variables during their modulation cycles
have not been studied in detail yet. Multicolour observations show that 
maximum magnitudes are bluer in the larger amplitude phase of the modulation. 
However, because of the lack of complete colour curves along the Blazhko 
cycle we still do not know whether these color changes are incidental to 
mean colour changes, or reflect only differences in the shape of the 
colour curves.  

The photoelectric observations of \citet{mitteil} also delimit us to
detect only $B-V$ colour changes of light maxima. Fig.~\ref{fold} shows 
the $B$, $V$ and $B-V$ magnitudes of maxima folded with the mean Blazhko 
period value (75.7~d). The $B$ and $V$ maximum magnitudes exhibit $0.2-0.3$~mag 
changes along the Blazhko period, while the amplitude of the $B-V$ colour of 
the maxima is about 0.06~mag, again, the colours of maxima are bluer in 
the larger amplitude phases of the Blazhko cycle.

As the sampling of the photoelectric data is not suitable for the investigation of
possible colour changes of the minima or the mean colours, the interpretation
of this result needs further multicolour observations of complete light 
curves along the entire Blazhko cycle.

\subsection{Radial velocity changes}

Three sets of radial velocity measurements of XZ Dra are available.
\citet{woolley} published 9 radial velocity observations that covered 
the different phases of the pulsation within 11 days (JD~$2\,438\,929
-2\,438\,940$). According to the B and V magnitudes of maxima, this 
interval fell within the minimal amplitude phase of the Blazhko period. 
The mean value of these radial velocity measurements is $\langle V_{\rm rad} 
\rangle =-30.83 \pm1.56$~km/s, as defined by fitting the actual pulsation 
frequency (determined from the light curve) and 2 of its harmonics to 
the radial velocity data. 

Radial velocity values obtained from spectra taken in 1971 with 
the 72 inch telescope of the Dominion Astrophysical Observatory, 
were published in \citet{mitteil}. These data were gathered during a 
longer time interval, and covered different phases of the Blazhko 
and the pulsation cycles as well. The typical error of these 
measurements was 5~km/s. Although radial velocity curves of Blazhko 
variables are different in the different phases of the Blazhko cycle,
its mean values during the pulsation cycles are supposed to remain 
the same. Therefore, it seems plausible to use all these measurements 
to determine the actual mean value of the radial velocity, instead 
of selecting observations belonging to a given Blazhko phase, 
as that would result in a much more limited sample with much more 
uncertain mean value. Even if all the radial velocity measurements 
from 1971 are used, the phase coverage of the pulsation is not complete 
enough to fit a reliable radial velocity curve when determining
its mean value. Adding 1-2 artificial data via polynomial interpolation 
helps the situation. The mean value of the so completed radial velocity 
data is determined similarly to that of Woolley's data (i.e, by using 
3rd order Fourier fit). As a result $\langle V_{\rm rad} \rangle 
=-32.0 \pm3.35$~km/s is obtained if uncertainty arising from the choice 
of the complementary data is also taken into account.

The third published data set of radial velocity observations of XZ Dra 
are the four measurements of \citet{layden} obtained around JD 2\,448\,131.
Although these observations show synchronous changes with the photometric 
variations, which exclude misidentification, their mean value $+77\pm 22$ km/s 
is very discrepant from the other two mean radial velocity values. Although 
the individual errors of Layden's measurements are very large, about 45~km/s, 
this cannot explain a systematic offset of about 100 km/s \citep{layden2001}.
We could not find any explanation for the deviation of these data, however, 
as no physical reason for such a large variation of the mean radial velocity 
seems to be plausible (see also Sect. 5.1), these data are regarded as unreliable.

As a summary we conclude in that the presently available radial velocity
measurements are not enough to draw any firm conclusion about the occurrence 
of real changes in the mean radial velocity of XZ Dra. To detect radial velocity 
changes connected to the $O-C$ variations, further extended accurate radial 
velocity observations are required.

\section{Possible explanations}

\subsection{Binary model}

Light time effect in a binary system is a plausible explanation for a 
periodic $O-C$ diagram. Binaries, however,  are extremely rare among 
RR Lyrae stars, and no indication of being a binary member has 
emerged previously for any Blazhko type RR Lyrae.

The binary orbit can be described by six parameters:
the semi-major axis $a$, the eccentricity $e$, the inclination $i$, 
the argument of periastron $\omega$, the epoch of periastron passage $\tau$, 
and the position angle of the line of ascending node $\Omega$.
The semi-major axis and the inclination are inseparable ($\xi=a\sin i$), 
and $\Omega$ cannot be determined at all in a non-eclipsing system.
 
Following \citet{Co71} we express the $O-C$ variation as a function 
of time with the orbital elements: 
\begin{equation}\label{phi}
O-C=c_0+{\xi(e^2-1)\over {c}}
\left[ {{\cos\omega\sin\nu}\over {1+e\cos\nu}}-
{{\sin\omega}\over {e(1+e\cos\nu)}} \right]\Bigg|^{\nu}_{\nu_0}
\end{equation}
where $c$ denotes the speed of light, $c_0$ is the correction term to take
into account the initial epoch from which the $O-C$ values have been calculated, 
$\nu=\nu(t)$ is the true anomaly (henceforward the notation of time 
dependence $t$ is omitted), and $\nu_0=\nu(t=\tau)$.

It is well-known that the two centre problem has no closed analytic
expression for $\nu(t)$. We use the classical formulae \citep[see
$e.g.$,][]{wint}:
\begin{equation}\label{anomaly}
\begin{split}
\sin\nu&=\\&2\sqrt{1-e^2}
\sum\limits_{k=1}^{\infty}J'_k(ke)\sin\left[k{{2\pi}\over 
{P_{\rm orb}}}(t-\tau)\right],\\
\cos\nu&=\\&-e+{{2(1-e^2)}\over {e}}\sum\limits_{k=1}^{\infty}J_k(ke)
\cos\left[k{{2\pi}\over
{P_{\rm orb}}}(t-\tau)\right], 
\end{split}
\end{equation}
where $P_{\rm orb}$, $J_k$ and $J'_k$ denote the orbital period of the binary, 
the Bessel function of first kind of order $k$, and its derivative, respectively,
and also the recursion relation for Bessel function  \citep[$e.g.$][]{as}: 
\begin{eqnarray}\label{Bessel}
J'_k(x)&=&{1\over 2}\left[J_{k-1}(x)-J_{k+1}(x)\right]. 
\end{eqnarray}

To find the model parameters of the supposed binary system, we need to solve 
the non-linear least-squares problem defined by Eq.~(\ref{phi}) and the 
$O-C$ data. The model fitting was carried out using the combined data set
of both individual and normal maximum timings (422 points). These data are 
shown in Fig.~\ref{oc} and Fig.~\ref{ocfold} by red and blue symbols, 
respectively. The data was first corrected for the continuous period increase 
according to Eq. (2). Weight factors were 2 and 1 for normal and individual 
maxima, respectively.
    
Substituting Eqs.~(\ref{anomaly}) into Eq.~(\ref{phi}) we obtain an
expression including six parameters ($\xi$, $e$, $\omega$, $\tau$, 
$P_{\rm orb}, c_0$). In practice, we solved the problem of parameter 
estimation by a combination of grid search method (using $e$ as control
parameter) and the Levenberg-Marquardt algorithm \citep{MarQ} by 
utilizing the facilities of the program package {\sc Mathematica}. 
Sums truncated to $30-100$ elements had been proven to have sufficient
accuracy and were calculated instead of infinite sums.

The parameter values obtained and the {\it rms} of the fits that characterize 
the goodness-of-fit are summarized in Table~\ref{BinFit}. We note that the 
time coverage of the $O-C$ data is much more complete than that of any 
similar data analyzed before.
The only RR Lyrae for which binary model solution has been calculated
is TU UMa, a non-Blazhko RRab star with $P_0=0.558$~d pulsation
period, its binary model \citep{Wade} is based only on 83 points. 
It is worth noting that both for XZ Dra and TU UMa, large eccentricity 
solutions have emerged ($e\approx 0.7 - 0.9$) and the orbital periods are 
also very similar, around 7100 and 8000 days, respectively.

\begin{table*}
\caption[]{Results of binary model fitting}\label{BinFit}
\begin{tabular}{@{\hspace{14pt}}l@{\hspace{26pt}}r@{\hspace{6pt}}l@{\hspace{68pt}}r@{\hspace{6pt}}l@{\hspace{68pt}}r@{\hspace{6pt}}l}
\hline
\noalign{\smallskip}
            Parameter     &\multicolumn{6}{c}{    Model} \\
\noalign{\smallskip}
\hline
\noalign{\smallskip}
 $e$                 &  \multicolumn{2}{l}{\hskip28pt 0.65} &
\multicolumn{2}{l}{\hskip28pt 0.8}
&\multicolumn{2}{l}{\hskip28pt 0.95} \\
 $P_{\rm orb}$[d]      &  $7113.8$&$\pm$ 51.2    & $7097.8$&$\pm$ 42.4   &
$7013.3$&$\pm$ 9.66  \\
 $\xi$ [AU]          &  $-2.443$&$\pm$ 0.145   & $-2.725$&$\pm$ 0.178  &
$-3.504$&$\pm$ 0.330 \\
 $\tau$ [JD]              &  $2427165.3$&$\pm$ 108.13 & $2427174.3$&$\pm$ 80.86
&
$2427334.9$&$\pm$ 16.93 \\
 $\omega$ [rad]      &  $1.2885$&$\pm$ 0.074   & $1.1531$&$\pm$ 0.082  &
$0.8778$&$\pm$ 0.097 \\
 $c_0$ [d]           &  $-0.01363$&$\pm$ 0.00111 &$-0.01319$&$\pm$ 0.00107
&$-0.01249$&$\pm$ 0.00102 \\
\noalign{\smallskip}
\hline
\noalign{\smallskip}
 $\sigma$ [d]    &\multicolumn{2}{l}{\hskip18pt 0.01404}&
\multicolumn{2}{l}{\hskip18pt 0.01397}&\multicolumn{2}{l}{\hskip18pt 0.01399} \\
\noalign{\smallskip}
\hline
\end{tabular}
\end{table*}

\begin{figure*}
\vskip -1cm
   \centering
   \includegraphics[width=18cm]{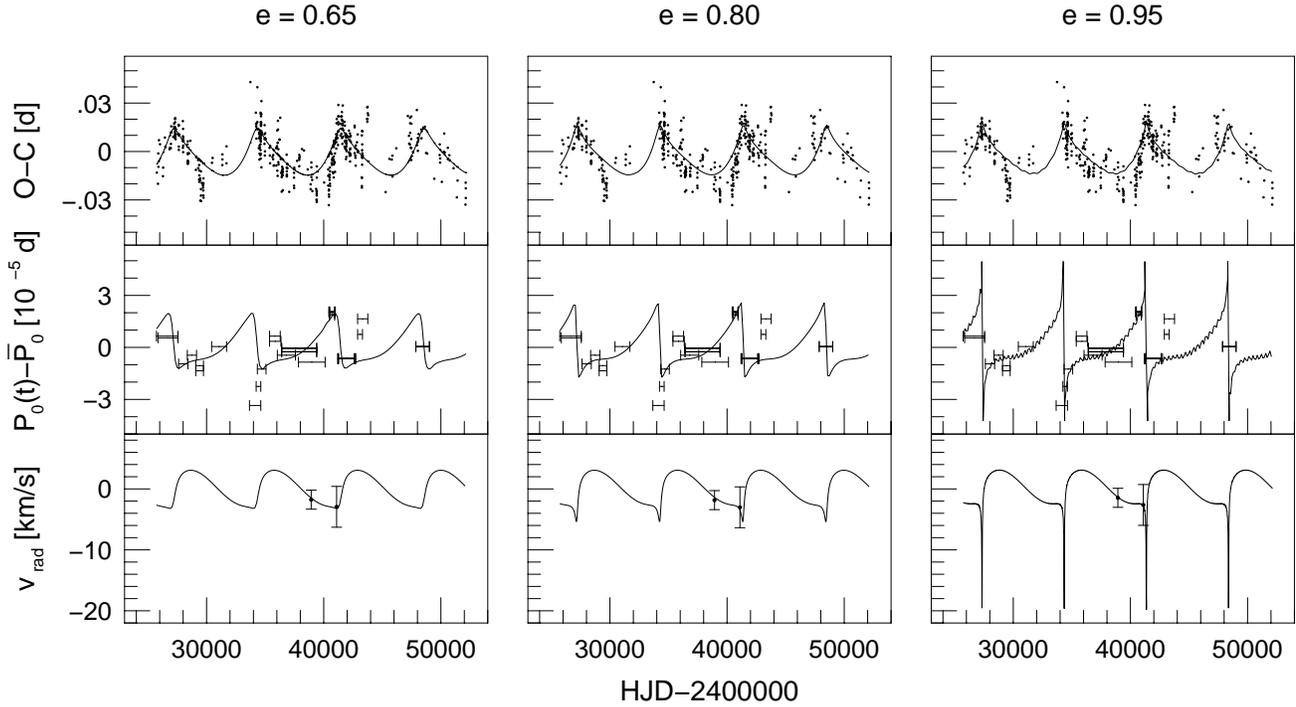}
   \caption{
Binary model solutions for the $O-C$ data assuming binary orbits of 0.65, 0.80,
and 0.95 eccentricities. Upper panels show the $O-C$ data over plotted by the 
different eccentricity binary model predictions. Middle and bottom panels show 
similar plots of the pulsation period and radial velocity data, respectively. 
Period differences using  $\overline{P_0} =0.4764955$ d mean pulsation period 
value are indicated and the time intervals that the observed periods correspond
to are also shown. Although global agreement between the observations and the 
corresponding fits is achieved, significant differences can be also seen.}
              \label{bin}
\end{figure*}

Although radial velocity data of XZ Dra are too few to make them useful
in  model-fitting, it is worth checking their values in comparison with
the binary model solutions. Predictions for possible radial velocity 
changes are also useful in order to obtain successful observational 
evidence pro or contra the binary hypothesis. To estimate the $\gamma$ 
velocity we minimized the least-square sum in the deviations of the measured 
mean radial velocities ($\langle V_{\rm rad} \rangle$, see Sect. 4.3) 
from the fits. The so obtained $\gamma$ velocities ($-29.06, -28.99$ and 
$-29.39$ km/s for the 0.65, 0.80 and 0.95 eccentricity solutions, respectively) 
were subtracted from the measured mean radial velocity values 
($v_{\rm rad(obs)}=\langle V_{\rm rad} \rangle-\gamma$) in order 
to compare them with the models.

The center-of-mass radial velocity can be calculated according to 
the formula:
\begin{equation}\label{Vrad}
v_{\rm rad}={{2\pi\xi}\over {P_{\rm orb}\sqrt{1-e^2}}}\left[
\cos\omega\left(\cos\nu+e\right)-\sin\omega\sin\nu\right].
\end{equation}
Substituting Eqs.~(\ref{anomaly}) into Eq.~(\ref{Vrad}) we arrive at
a formula that expresses the radial velocity of the center-of-mass as 
a function of time. 

Fig.~\ref{bin} compares the observed $O-C$, pulsation period
($P_0(t)-\overline{P_0}; \,\, \overline{P_0} =0.4764955$ d) and 
center-of-mass radial velocity values with predictions of three 
possible binary models listed in Table~\ref{BinFit}.

The three different eccentricity models shown are in similarly good agreement
with the observations. However, both the $O-C$ and period data indicate 
significant differences as well. None of the models follows the very 
steep increase of the $O-C$ around JD 2\,440\,500, and there are also 
outlying observed period values according to each of the solutions. 

The overall fitting accuracy of the $O-C$ is 0.014 d ($\approx$ 20 min) 
which also seems to be larger than expected. Although the Blazhko modulation
accounts for some real scatter of the $O-C$ data, data analysis 
indicates that the $O-C$ variation during the Blazhko period does 
not have amplitude larger than $0.005-0.010$~d.

We can thus conclude that, although, the data can be globally described
within the frame of a reliable binary model, period changes generated by 
other mechanisms do also occur. This is, however, not surprising as 
random period changes in RR Lyrae stars are common. Conclusive evidence 
of the binary model would be the detection of real center-of-mass radial 
velocity variation during the 7200 d cycle. The binary model predicts 
minimal and maximal values for the center-of-mass velocity to occur 
in 2011 and during the $2012-2018$ period, respectively.

Another serious  deficiency of the binary model is that it cannot give an 
explanation for the simultaneous changes detected in the Blazhko period. 
The connected changes of the pulsation and Blazhko periods indicate 
common physical background. The light-time effect results in Blazhko period 
changes as well, however, this is far below the detectability limit. 
It would give rise to Blazhko period changes of the order of $10^{-3}$~d, 
which is 3 orders below what is actually detected.
 
\subsection{Magnetic cycles}

The question of magnetic field in Blazhko stars is still a matter of 
debate from the observational point of view. However, the obliquely 
oscillating magnetic rotator model is one of the possible explanations 
for the phenomenon \citep{cousens,shiba}.

Thanks to recent technical developments we have been able to learn more and 
more about the magnetic field structure and the $11-22$-year magnetic cycle 
of the Sun. It is already well established that, as a consequence of  
changes in the global field strength and structure during the magnetic cycle, 
slight changes in most of the global solar parameters and also in both the 
{\it p} and {\it f} mode oscillationary frequencies are taking place
\citep[see e.g.,][]{dziem,li,zieba}.

The existence of long-term cycles in late type active stars, 
which manifests itself in changes of the activity level of these stars,
probably indicates also cyclic variations in their global magnetic fields
\citep[and references therein]{olah}.

Although we do not have evidence yet of magnetic cycles in evolved stars
we cannot exclude its possibility. In the framework of the oblique magnetic 
rotator model of the Blazhko phenomenon any long-term cyclic behaviour 
seems to be a natural consequence of changes in the global magnetic field
strength. Thus the four-year cycle of RR Lyrae may also indicate a similar
explanation \citep{detre}. According to recent results \citep{jbsz} the 
light curves of RR Lyrae correspond to that of non-Blazhko RRab stars 
during the minimal phase of the four-year cycle, when the amplitude of 
the modulation is small. This suggests, that if the oblique magnetic 
rotator-pulsator model, which describes the observed modulation 
of the light curve is valid, then the diminution of the modulation indicates 
weakening in the global magnetic field strengths, consequently resulting in 
normal type, undisturbed light changes.

The long-term modulations of RR Lyrae and XZ Dra diverge, however,
practically in all aspects of their observed properties, as summarized in
Table~\ref{comp}. This cautions us, that, if the magnetic cycle hypothesis
is valid for these stars, its detailed nature and physic should be very 
complex resulting in different phenomena in different stars. 

\begin{table}[!bbbb]
\caption[]{Comparison of the long-term cyclic behavior of XZ Dra and RR Lyr 
}\label{comp}
\begin{tabular}{lcc}
\hline
\noalign{\smallskip}
      &  XZ Dra & RR Lyr\\
\noalign{\smallskip}
\hline
\noalign{\smallskip}
cycle length & ~20 years & ~4 years\\
\noalign{\smallskip}
\hline
\noalign{\smallskip}
\multicolumn{3}{c}{ detected changes along the long-term cycle}\\
\noalign{\smallskip}
\hline
\noalign{\smallskip}
 changes in $P_0$& yes & no\\
 changes in $P_{\rm B}$& yes & no\\
phase shift in $P_{\rm B}$& no & yes\\
amplitude changes of the modulation& no&yes\\
\noalign{\smallskip}
\hline
\end{tabular}
\end{table}

The pulsation period of XZ Dra varies by about $0.00004-0.00005$~d. 
Very small changes of the global stellar parameters (radius, effective 
temperature, luminosity) can induce such a period change, that 
may occur as a consequence of changes in the strength of the global 
magnetic field, i.e, a magnetic cycle. \citet{stoth} calculated period 
changes caused by hydromagnetic effects in RR Lyrae stars and concluded 
that the observed period changes do not contradict such an explanation.
Recent results for the Sun show that during the solar cycle structural 
adjustments in the solar interior are taking place which induce detectable 0.1\% 
photospheric temperature and 0.02\% total irradiance changes. The radius 
changes of the Sun are estimated to be of the order of 0.001\% \citep{li}. 
These detected changes of the solar parameters make it feasible that 
during a magnetic cycle changes in the global parameters of an RR Lyrae 
star may also occur. 

This picture does not exclude the possibility of the connected changes in
the Blazhko period (that is the same as the rotational period according to
this model) either, but to estimate its rate and sign would only be possible 
with the help of a quantitative description of the model.
To work out such a model in details is, however, far beyond the scope of the
present paper, nevertheless should be the subject of further theoretical works.

\begin{figure}
   \centering
   \includegraphics[width=7.5cm]{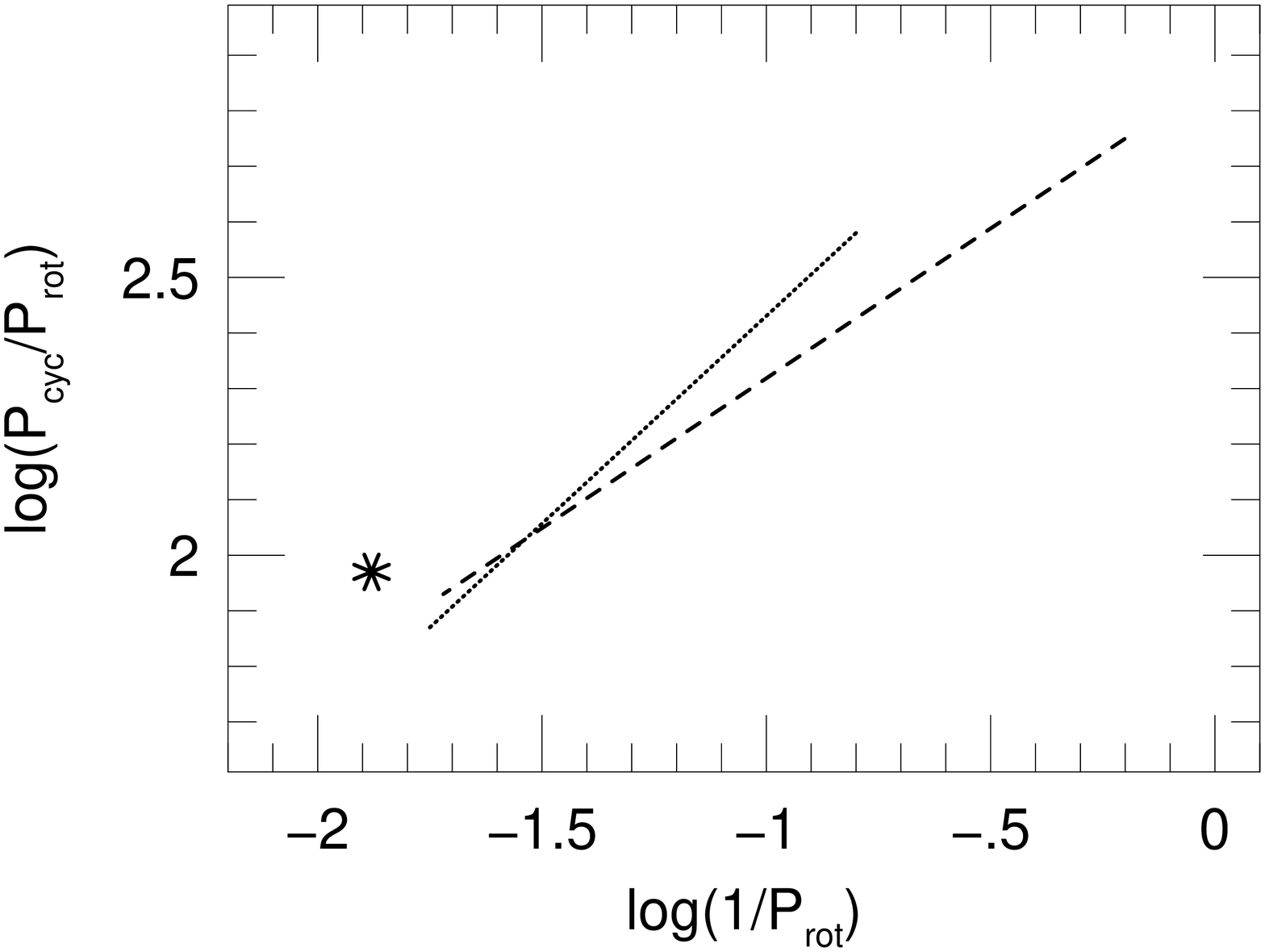}
   \caption{
The $\log(P_{\rm cycle}/P_{\rm rot}) \, vs. \,  \log(P_{\rm rot})$
connection
that characterizes the stellar dynamo. Dotted line indicates
the location of lower main-sequence stars which activity cycles were
determined from the variations of their  CaII fluxes \citep{bal}.
Dashed line shows the same relation found by \citet{ok}
for active stars which cycle lengths have been reliably determined from
photometric measurements. Star denotes the location of XZ Dra using 
the 76 d Blazhko period for $P_{\rm rot}$ and 7200 d for $P_{\rm cycle}$.     
}
              \label{rot}
\end{figure}

Assuming that the 76-day Blazhko modulation of XZ Dra reflects
its rotation rate as proposed by the oblique magnetic rotator-pulsator
model, and that the 7200-day variation of the pulsation and Blazhko 
periods are connected to cyclic variations of the global magnetic field,
the resultant $P_{\rm cyc}/P_{\rm rot} \,\, vs. \,\, 1/P_{\rm rot}$ 
quantities serve as an observational check for any possible dynamo 
mechanism following the conception outlined in e.g. \citet{bal}. 
Fig.~\ref{rot} documents that these two quantities for XZ Dra are 
in excellent agreement with results obtained both for lower 
main-sequence stars from CaII fluxes \citep{bal}
and for the complete sample of stars showing cyclic behaviour 
according to their photometric properties \citep{ok}.
This result indicates that the magnetic cycle explanation for the 7200 d 
periodicity of XZ Dra seems to be reliable.

\section{Concluding remarks}

The 70 years photometric observations of XZ Dra have revealed that long 
period (7200 d) cyclic changes in the pulsation period have been 
occurring. The Blazhko period seems to follow this period change,
exhibiting $3-5$ days full range of period change. This means that, 
besides RR Lyrae, XZ Dra is the second Blazhko variable which clearly 
shows indication of long-term cyclic behaviour. The manifestation of 
these long-term changes is, however, completely different for the 
two stars.

The long-term cyclic changes favour the magnetic rotator-pulsator
model of the Blazhko modulation, by explaining the observed phenomena 
with changes in the global magnetic field structure and/or strength.
However, to check the reality of this explanation, detailed 
theoretical work is needed. 

Binary interpretation of the observations, although giving acceptable
good fit to the data, fails to explain the detected range of the Blazhko 
period variation.

Rapid $O-C$ and radial velocity changes of XZ Dra are predicted to occur
next time in the years $2011-2018$, when coordinated photometric and 
spectroscopic observations would greatly help to give correct answers 
to the presently unsolved questions.

\begin{acknowledgements}
We thank Andrew Layden for sending us his radial velocity measurements.
Thanks are also due to Andrew Wilkins for correcting the language of the paper.
This research has made use of the SIMBAD database, operated at CDS 
Strasbourg, France. This work has been supported by OTKA grants T30954 and 
T30955.

\end{acknowledgements}
\bibliographystyle{aa}

\end{document}